\documentclass[amsmat,amssymb,amsfonts,aps,prb,twocolumn,showpacs]{revtex4}

\usepackage{graphicx}

\usepackage{xcolor}

\begin{document}

\title{Electronic transport of folded graphene nanoribbons}


\author{Jhon W. Gonz\'alez}
\address{International Iberian Nanotechnology Laboratory, 
Av. Mestre Jos\'{e} Veiga, 4715-330, Braga, Portugal, and
\\Departamento de F\'{i}sica, Universidad T\'{e}cnica Federico Santa Mar\'{\i}a, Casilla postal 110 V, Valpara\'{i}so, Chile}

\author{M\'onica Pacheco}
\address{Departamento de F\'{i}sica, Universidad T\'{e}cnica Federico Santa Mar\'{\i}a, Casilla postal 110 V, Valpara\'{i}so, Chile}
\author{Pedro Orellana}
\address{Departamento de F\'{i}sica, Facultad de Ciencias,  Universidad Cat\'olica del Norte, Casilla postal 1280, Antofagasta, Chile}
\author{Luis Brey}
\author{Leonor Chico}
\address{Instituto de Ciencia de Materiales de Madrid, Consejo Superior de
Investigaciones Cient{\'{\i}}ficas, Cantoblanco, 28049 Madrid, Spain }

\begin{abstract}

We investigate the electronic transport properties of a folded graphene nanoribbon with 
monolayer nanoribbon contacts. We consider two possible foldings: either 
the nanoribbon can be folded onto itself in the shape of a hairpin with the nanoribbon leads at a $0^\circ$ angle, 
or the monolayer contacts have different directions, forming a $60^\circ$ angle.
The system is described by a single $\pi$-band nearest-neighbor tight-binding Hamiltonian taking into account curvature effects.
We have found that for the case of a nanoribbon folded over 
itself the conductance oscillates from almost zero and a finite value 
depending on the coupling between contacts, whereas in the $60^\circ$ angle folding the conductance is only slightly perturbed, 
allowing for the connection of graphene nanoelectronic components in a variety of geometries.

\end{abstract}

\maketitle

\section{Introduction}
Graphene is an one-atom-thick
covalently-bonded carbon layer ordered in a honeycomb lattice. 
It has been recently isolated and shown to be stable under room 
conditions \cite{Novoselov_2004}, giving rise to the exploration of its fascinating properties. 
As it is an ambipolar material,  the charge carrier type can be changed 
 by applying a gate voltage; furthermore, the high mobility of charge carriers \cite{Neto_RMP} due 
to the absence of backscattering. These characteristics point at graphene as an interesting alternative for novel nanoelectronic devices. 
However, as graphene is a zero-gap material, it cannot be employed for the fabrication of diodes and most electronic 
components, for which an electronic gap is essential. Thus, 
there has been substantial interest in controlling the electronic properties of graphene by quantum size effects.  
For example, cutting graphene into quasi-one-dimensional nanoribbons \cite{Han_PRL_98} produces electronic gaps. 
Another route to produce a gap is to stack two graphene layers. Bilayer graphene  \cite{Neto_PRL99,Nilsson_PRB76,Min_PRB75,jw_PRB}  holds huge 
potential because it is a  gate-tuneable semiconductor, \cite{Guinea_PRB73,McCann_PRL96} as observed  in electronic transport 
\cite{Neto_PRL99,Gava_PRB79} and in photoemission experiments \cite{Ohta_Sci313}.

Due to the planar geometry of graphene, large-scale integrated circuits made by nanolitographic techniques can 
 be envisioned. Indeed, there are numerous studies on the transport properties of graphene nanoribbon bends and junctions 
 such as those that may occur in graphene-based nanoelectronics. For certain angles and ribbon geometries, the conductance can be completely
 suppressed, and in certain cases,  conductance oscillations due to multiple resonances may appear \cite{Brey_PRB_78}.
 
 In order to avoid undesirable conductance gaps, we propose to use folded ribbons as contacts. 
   On the one hand, it allows integration in the vertical direction, since the ribbon can connect stacked graphene components. On the other hand, using a sufficiently long ribbon, it can serve as a connector at different angles without a conductance gap.



In this work we present a theoretical study of the conductance through folded metallic armchair graphene nanoribbons (aGNR). 
Such fold terminations for bilayer graphene have been experimentally observed: studies of heat treated graphite 
using high resolution TEM 
images show a predominance of closed edges between layers of graphene and a large proportion of AA 
stacking \cite{Iijima_PRL102}. 
Furthermore, folds and bends on narrow 
carbon nanoribbons have also been reported \cite{Li_Science319}. From the theoretical viewpoint, the electronic structure of folded graphene has been studied \cite{Okada_2010}.  The transport properties of folded 
graphene ribbons {\it along } the fold direction have also been reported without considering curvature effects \cite{Xie_2009}; besides, the transport in a uniform magnetic field has also been explored \cite{Elsa_pp}, but the emphasis
was in the effective change of direction the magnetic flux when moving across the fold, and no curvature effects nor interlayer coupling
were considered. 

\begin{figure}
  \centering
\includegraphics[clip,width=8.cm,angle=0]{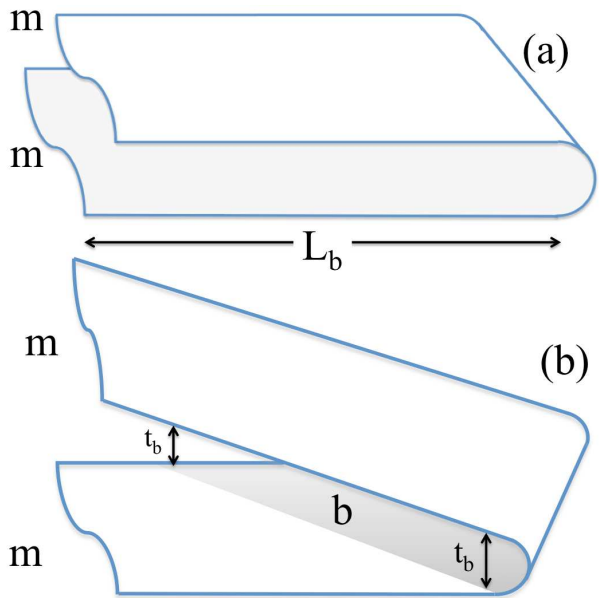}
\caption{Schematic view of the two configurations considered. (a) Nanoribbon folded at $0^\circ$. The length of the bilayer part is $L_b$. To this bilayer portion, two monolayer nanoribbons are connected, indicated by $m$. (b)  
Geometry of the $60^\circ$ folded ribbon. The bilayer region is a triangular portion marked with $b$; the monolayer contacts are labeled by $m$. Note that the distance $t_b$ between the upper and lower ribbon is constant in the planar part.}
\label{fig1:esq}
\end{figure}

We consider the folded ribbon to be composed of a curved portion, also called fractional nanotube, \cite{Huang_PNAS_09,Feng_PRB_80}
 and two armchair graphene ribbons with a finite overlap region, as depicted schematically in Fig. \ref{fig1:esq}. 
We explore the role of curvature and interlayer coupling on the conductance, showing that 
folded ribbons can be employed as metallic contacts provided that the effective bilayer region is minimized. 
The geometries studied are the following: 
(i) a nanoribbon folded onto itself, in the shape of a hairpin, so that the electrode ribbons are on top of each other, oriented at a 
$0^\circ$ angle, shown in Fig. \ref{fig1:esq}(a); (ii) the ribbon is folded in such a way that the electrode ribbons are at a $60^\circ$ angle, depicted in Fig. \ref{fig1:esq}(b). We choose this
angle in order to achieve an AA stacking in the bilayer region formed by the overlapping electrodes. Other angles can 
be considered with a different bilayer stacking, but for the sake of simplicity we restrict ourselves to the most symmetric case
which has been recently observed to experimentally appear in bilayer graphene \cite{Iijima_PRL102}.



\section{\label{sec:Model} Theoretical description of the system}

\subsection{Tight-binding Hamiltonian}

\begin{figure}
  \centering
\includegraphics[clip,width=8.2cm,angle=0]{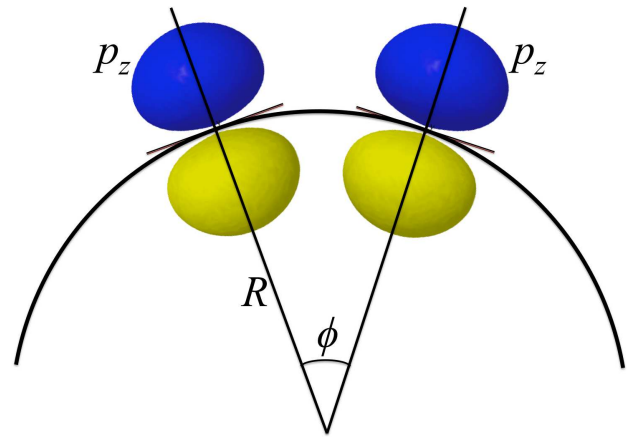}
\caption{Schematic drawing of two misaligned $p_z$ orbitals over a carbon nanotube, where the misalignment angle $\phi$
is indicated.
}
\label{orbit}
\end{figure}

The low-energy transport properties of planar graphene are well described within a single $\pi$-band tight-binding 
approximation with nearest-neighbor in-plane hopping given by the parameter $\gamma _0 \approx$ 3 eV.
This model only considers the  $\pi$ and $\pi^\ast$ bands due to the hopping between $p_z$ orbitals perpendicular to the
graphene plane. 
In the fractional nanotube region, we must consider the effect of curvature. As its main consequence is the misalignment of the 
 $p_z$ orbitals, we model it by assuming that the hopping integral between neighbor atoms is changed with respect to that of planar graphene, 
 The curvature-modified  hopping $\tilde{\gamma _0}$ is proportional to the cosine of
 the misalignment angle $\phi$ between orbitals,  $\tilde{\gamma _0}=\gamma_0 \cos \phi$, with $\phi$ defined in Fig. \ref{orbit}  \cite{Kleiner_PRB63}. 
 
%
%
%
%
%
The region where the two ribbons overlap is just a finite-size fragment of bilayer graphene with AA stacking. 
In the case of folding at $0^\circ$, the bilayer region is a rectangular flake of length $L_b$. For nanoribbons folded at 
 $60^\circ$, the bilayer flake has a triangular shape, as depicted in Fig. \ref{fig1:esq} (a), 
with size depending on the ribbon width. 
The interlayer coupling is modeled with a single hopping $\gamma_1$ connecting 
atoms directly on top of each other, which we take as
$\gamma_1 = 0.1\gamma_0$, in agreement with experimental results \cite{Ohta_Sci313,Malard_PRB2007}.

In the bilayer with AA stacking all the atoms of layer $1$ are on top of the
equivalent atoms of layer $2$. We assume that all atoms 
bottom layer are connected to those on the 
the upper layer located exactly on top of them; 
thus, the Hamiltonian takes the form 
\begin{eqnarray}
H^{AA}=& - & \gamma _0 \sum _{<i,j>,m} (a^+_{m,i} b_{m,j} +
h.c.)
\nonumber \\
& - &\gamma_1 \sum _{i} (a^+_{1,i} a_{2,i} + b^+_{1,i} b_{2,i}+h.c.). 
\label{H_TB_AA}
\end{eqnarray}
where $a_{m,i} (b_{m,i})$ annihilates an electron on sublattice $A
(B)$, in plane $m=1,2$, at lattice site $i$. The subscript $<i,j>$
represents a pair of in-plane nearest neighbors. The second term in Eq. (\ref{H_TB_AA}) 
represents the interlayer hopping.

\subsection{Landauer-B\"uttiker formalism}
We calculate the electronic and transport
properties using the surface Green function matching method \cite{Chico_1996b,Nardelli_1999}. 
Thus, the system is
partitioned in three blocks: two semi-infinite leads, which we
assume to be semi-infinite aGNRs, and the conductor, consisting of the
bilayer and the folded region. The Hamiltonian of the system can be partitioned as
\begin{equation}
H= H_C  + H_R + H_L + V_{LC} + V_{RC},
\end{equation}
where $H_C$, $H_L$, and $H_R$ are the Hamiltonians of the central
portion, left and right leads respectively, and $V_{LC}$, $V_{RC}$
are the coupling matrix elements from the left $L$ and right $R$
lead to the central region $C$. The Green function of the conductor
is
\begin{equation}
\mathcal{G}_C(E) = (E-H_C - \Sigma_L -\Sigma_R)^{-1},
\end{equation}
where $\Sigma_\ell= V_{\ell C}g_\ell V_{\ell C}^\dagger$ is the
selfenergy due to lead $\ell=L,R$, and $g_\ell = (E -H_\ell)^{-1}$
is the Green function of the semi-infinite lead $\ell$ \cite{jw_PhysicaB404}.

The conductance can be calculated
within the Landauer formalism in
terms of the Green function of the system\cite{Datta_book,jw_EPL}:
\begin{equation}\label{LandauerG}
G = \frac{{2e^2 }}{h}T\left( {E } \right) = \frac{{2e^2 }}{h}
{\mathop{\rm Tr}\nolimits} \left[ {\Gamma _L G_C \Gamma _R
G_C^\dagger} \right],
\end{equation}
where  $T\left( {E } \right)$, is the transmission function across
the conductor, and  $\Gamma_{\ell}=i[ {\Sigma _{\ell}  - \Sigma
_{\ell} ^{\dag} }]$ is the coupling between the conductor and the
$\ell=L,R$ lead.

\section{\label{sec:results} Results}

\subsection{Nanoribbons folded at $0^\circ$}

We first focus on 	
an armchair graphene nanoribbon 
folded onto itself in the shape of a hairpin, i.e., 
in a configuration where the relative angle between the bottom 
and top leads is  
$0^\circ$ (Fig.\ref{fig1:esq}(a)).  
Here we need to consider two effects: the layer-layer coupling which depends on the length of the 
overlapping region, and the effects of the curved portion, modeled as a fractional nanotube. We assume our leads 
are noninteracting monolayer graphene nanoribbons, so that the bilayer region with interlayer coupling has a finite size $L_b$, given in terms of the 
armchair unit cell size length. The width of the nanoribbon is given by the number of dimers across its width, N.
%
%

\begin{figure}
  \centering
\includegraphics[width=\columnwidth,clip]{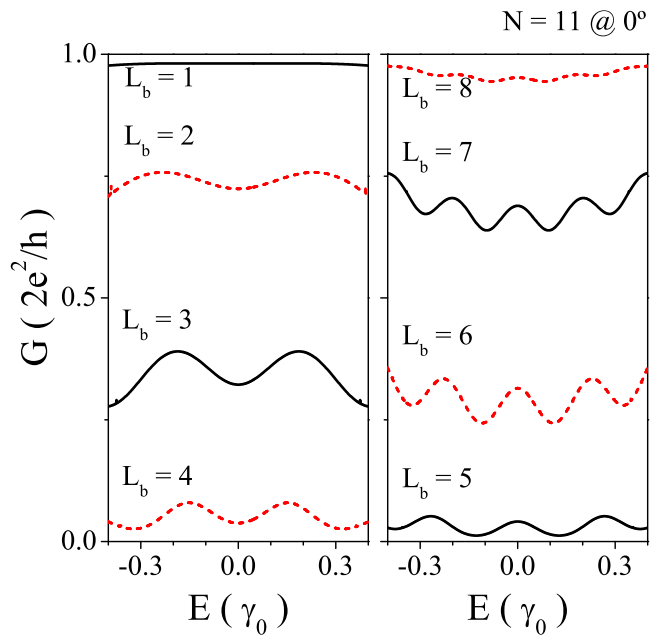}
\caption{(color online) Conductance as a function of Fermi energy for an aGNR of width N = 11 at $0^\circ$ with a 
fractional (4,4) armchair carbon nanotube for several bilayer lengths $L_b$,
showing a full period as a function of the overlapping size. 
The label $L_b$ is included in the graph.
}
\label{fig2:N11CN44}
\end{figure}

For an armchair nanoribbon folded at $0^\circ$, the curved edge has to be a fractional armchair nanotube (\textit{f}aCNT).
 In this work we model the curved edge as a section of a (4,4) armchair carbon nanotube.
%
In Figure \ref{fig2:N11CN44} we show the conductance of the aGNR of width N = 11 folded at 
$0^\circ$ as a function of Fermi energy for several bilayer lengths $L_b$.
Notice the periodic behavior in the conductance as a function of the energy, due to the multiple interferences 
produced in the bilayer region. These periods change with the system size. Besides, there is another periodicity related to the bilayer portion $L_b$, roughly 
equal to 8 unit cells. 
These periodicities in the electronic conductance are also observed  
with equal frequencies in wider aGNR ribbons folded at $0^\circ$, such as the N=17 case (not shown here because in the depicted energy range their conductances are equal, given that they only have one conductance channel at that energy interval). 
The oscillations are reminiscent of those found in the transmission through bilayer flakes \cite{jw_PRB}, where the conductance was demonstrated to depend on both the energy and the size of the scattering region.

\begin{figure}
  \centering
\includegraphics[width=\columnwidth,clip]{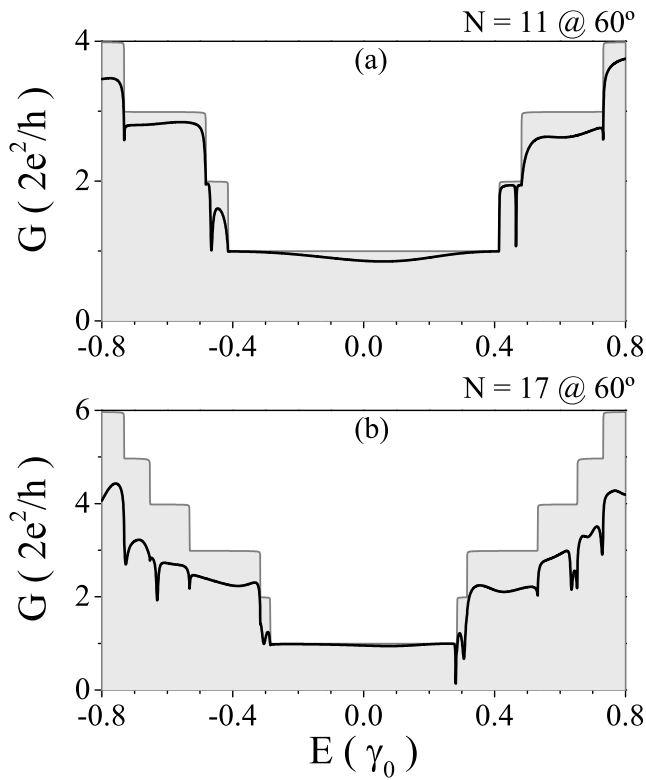}
\caption{(color online) (a)  Conductance as a function of the Fermi energy for an aGNR of width N = 11 at $60^\circ$ with a 
fractional zigzag carbon nanotube (8,0). 
(b) Conductance as a function of the Fermi energy for an aGNR of width N = 17 at $60^\circ$ with a 
fractional zigzag carbon nanotube (9,0). 
For reference we included a dotted line for the respective pristine monolayer aGNRs, namely, the N = 11 and the N = 17 cases.
}
\label{fig5:N11CN80}
\end{figure}

\subsection{Nanoribbons folded at  $60^\circ$ }

For the $60^\circ$ configuration, the folded edge for an armchair nanoribbon has to be a fractional  zigzag nanotube (\textit{f}zCNT) 
in order to maintain the 
AA stacking in the bilayer region. 
The radius of the \textit{f}zCNT depends on  nanoribbon width, 
and  close to $0.35$ nm, to be consistent with previous ab-initio 
calculations \cite{Feng_PRB_80}.

In figure \ref{fig5:N11CN80} (a), we show the conductance as a function of Fermi energy for aGNR N = 11 at 
$60^\circ$ with a \textit{f}zCNT (8,0). The bilayer region for this case consists of a triangle, as shown in Fig.\ref{fig1:esq}(b). 
The most relevant change with respect to to hairpin geometry is that the electron-hole 
symmetry is broken. 
This is due to the mixing of the two sublattices produced by the interlayer hopping: when the two nanoribbon electrodes are
at $60^\circ$, the hopping in the bilayer region produces odd-numbered carbon rings comprising atoms of the bilayer and the 
fractional nanotube region. This did not happen when the contacts were at  $0^\circ$, thus yielding e-h symmetric conductances, like  
those previously shown in Fig. \ref{fig2:N11CN44}.

In figure \ref{fig5:N11CN80} (b) we show the conductance as a function of Fermi energy for an aGNR of width N = 17 at 
$60^\circ$ with a \textit{f}zCNT (9,0). Due to the size of the overlapping ribbons, we need to consider here a slightly larger nanotube than in the 
previous case. The chosen  \textit{f}zCNT is consistent with ab-initio predictions and permits 
the obtention of the AA stacking in the bilayer region. We also observe in this case that electron-hole symmetry is broken, 
but the decrease in in the electronic conductance around zero is very small, as in the N=11 aGNR folded at $60^\circ$ ribbons. 

Obviously, the folded nanoribbons always should show a conductance smaller than the pristine unfolded case, as can be seen in Figs. \ref{fig5:N11CN80} (a) and (b). This is due to the interferences occurring in the scattering folded region, both at the bilayer stacked portions and in the curved nanotube, where there is a different effective hopping because of curvature effects. But due to the small size of the scattering regions in the nanoribbons folded  at $60^\circ$, the change in the conductance around zero energy is not dramatic, thus allowing for the connection of nanoribbons at different heights without a substantial decrease in their transport properties.


%
%

\section{\label{sec:Conclusions} Conclusions}
We have studied the transport properties in folded graphene nanoribbons. We have found 
that the connections with folded ribbons do not show the large conductance gaps present in other types on nanoribbon 
junctions. We have also explained the lack of electron-hole symmetry present in folded graphene edges \cite{Okada_2010}, 
which can be understood in terms of the introduction of odd-membered carbon rings in the graphene lattice.
First-principles calculations show that the curved edge may have a larger radius \cite{Feng_PRB80}, so 
our approach overestimates the reduction of the conductance due to the curvature-induced 
change of the hopping integrals, pointing at an optimal conductance of folded ribbons as carbon connections.

\section*{Acknowledgments}
This work has been partially supported by the Spanish DGES under
grant FIS2009-08744 and the CSIC-CONICYT grant 2009CL0054.
J.W.G. would like to gratefully acknowledge helpful discussion to Dr. L. Rosales, 
to the ICMM-CSIC for hospitality and MECESUP research internship program. 
L. C. and J. W. G. acknowledge helpful discussions with A. Ayuela and the hospitality of the DIPC. 
M.P. acknowledges the financial support of FONDECYT 1100672 and DGIP/USM 11.11.62 internal grant.
P. O.  acknowledges the financial support of FONDECYT grant 1100560.


\end{document}